\documentclass[a4]{scrartcl}

\usepackage{tikz}
\usepackage{xspace}
\usepackage[hidelinks]{hyperref}
\usepackage{amsmath,amssymb,amsthm}
\usepackage{csquotes}
\usepackage{booktabs}
\usepackage{enumitem}
\usepackage[nameinlink,capitalize,noabbrev]{cleveref}
\usepackage[left=1.9cm,right=1.9cm,top=1.4cm,bottom=2.5cm]{geometry}
\usepackage[
    backend=biber,
    style=numeric-comp,
    minnames=5,
    maxnames=5,
    backref=true
]{biblatex}
\newcommand{\ttwo}{\textnormal{T2W}\xspace}
\newcommand{\tone}{\textnormal{T1W}\xspace}
\newcommand{\tonec}{\textnormal{T1CE}\xspace}
\newcommand{\flair}{\textnormal{FLAIR}\xspace}
\newcommand{\brainclusteringnospace}{BrainClustering}
\newcommand{\brainclustering}{BrainClustering\xspace}
\newcommand{\pixtopix}{Pix2Pix\xspace}

\newcommand{\unet}{UNet}
\newcommand{\resnet}{ResNet}
\newcommand{\unetone}{\unet-128\xspace}
\newcommand{\unettwo}{\unet-256\xspace}
\newcommand{\nifti}{NIfTI\xspace}
\newcommand{\pixtopixhd}{Pix2PixHD\xspace}
\newcommand{\brats}{BraTS\xspace}

\newcommand{\missingref}[1]{~[{\color{red}Missing Ref}]}

\newcommand{\kmeansoned}{$ k $-means1d\xspace}
\newcommand{\complementary}{Complementary\xspace}
\newcommand{\randomapproach}{Random\xspace}
\newcommand{\deepmedic}{DeepMedic\xspace}
\newcommand{\threed}{three-dimensional\xspace}
\newcommand{\twod}{two-dimensional\xspace}

\newcommand{\expe}[2]{\ensuremath{\mathbb{E}_{#2}\left[{#1}\right]}}
\newcommand{\traintest}{Train\&Test\xspace}
\newcommand{\search}{Search\xspace}
\newcommand{\patient}[1]{\texttt{#1}\xspace}
\newcommand{\tonelong}{T1-weighted\xspace}
\newcommand{\ttwolong}{T2-weighted\xspace}

\newcommand{\train}{\texttt{training}\xspace}
\newcommand{\validation}{\texttt{validation}\xspace}
\newcommand{\testuk}{\texttt{testOUR}\xspace}
\newcommand{\testbrats}{\texttt{test\brats}\xspace}
\newcommand{\testnotruth}{\texttt{testNoTruth}\xspace}

\newcommand{\email}[1]{\href{mailto:#1}{#1}}
\newcommand{\etal}{~{et al.}}
\definecolor{python_red}{HTML}{FF0000}
\definecolor{python_green}{HTML}{008000}
\definecolor{python_blue}{HTML}{0000FF}
\definecolor{python_gold}{HTML}{FFD700}
\definecolor{python_lightblue}{HTML}{00BFFF}
\definecolor{python_magenta}{HTML}{DDA0DD}
\definecolor{python_maroon}{HTML}{800000}
\definecolor{python_gray}{HTML}{808080}

\definecolor{darkorange}{HTML}{d95f02}
\definecolor{lightorange}{HTML}{fc8d62}

\definecolor{darkturquoise}{HTML}{1b9e77}
\definecolor{lightturquoise}{HTML}{66c2a5}

\definecolor{darkpurple}{HTML}{7570b3}
\definecolor{lightpurple}{HTML}{8da0cb}

\definecolor{darkpink}{HTML}{e7298a}
\definecolor{lightpink}{HTML}{e78ac3}

\definecolor{darkgreen}{HTML}{66a61e}
\definecolor{lightgreen}{HTML}{a6d854}

\definecolor{darkyellow}{HTML}{e6ab02}
\definecolor{lightyellow}{HTML}{ffd92f}

\definecolor{darksoil}{HTML}{a6761d}
\definecolor{lightsoil}{HTML}{e5c494}

\definecolor{tumorblue}{HTML}{a9cee5}
\definecolor{tumorpink}{HTML}{fbb4ae}
\definecolor{tumoryellow}{HTML}{ffff99}

\definecolor{redt}{HTML}{e41a1c}
\definecolor{lredt}{HTML}{fbb4ae}
\definecolor{bluet}{HTML}{377eb8}
\definecolor{lbluet}{HTML}{b3cde3}
\definecolor{greent}{HTML}{4daf4a}
\definecolor{lgreent}{HTML}{ccebc5}
\definecolor{purplet}{HTML}{984ea3}
\definecolor{lpurplet}{HTML}{decbe4}
\definecolor{oranget}{HTML}{ff7f00}
\definecolor{loranget}{HTML}{fed9a6}

\definecolor{original}{HTML}{003c30}
\definecolor{random}{HTML}{feb24c}
\definecolor{comp}{HTML}{525252}

\definecolor{bccolorone}{HTML}{b10026} % Red
\definecolor{bcfivecolorone}{HTML}{d95f02} % Orange
\definecolor{bctencolorone}{HTML}{e7298a} % Pink

\definecolor{bccoloronemed}{HTML}{ae4b60} % Red
\definecolor{bcfivecoloronemed}{HTML}{fc8d62} % Orange
\definecolor{bctencoloronemed}{HTML}{e78ac3} % Pink

\definecolor{pixcolorone}{HTML}{084594} % Blue
\definecolor{pixcolortwo}{HTML}{31a354} % Green
\definecolor{pixcolorthree}{HTML}{7570b3} % Purple
\definecolor{pixcolorfour}{HTML}{1c9099} % Teal

\definecolor{pixcoloronemed}{HTML}{467abe} % Blue
\definecolor{pixcolortwomed}{HTML}{6fc188} % Green
\definecolor{pixcolorthreemed}{HTML}{8da0cb} % Purple
\definecolor{pixcolorfourmed}{HTML}{55b0b7} % Teal

\newcommand{\nogtdataset}{\texttt{no-ground-truth}\xspace}
\newcommand{\hausdorff}{undirected $ 95^{\text{th}} $ Hausdorff distance\xspace}

\begin{document}

\title{MRI Scan Synthesis Methods based on Clustering and \pixtopix}

\author{Giulia Baldini\thanks{This work was partly funded by DFG, project numbers 416767905 and 456558332.}~\thanks{G. Baldini (\email{giulia.baldini@hhu.de}) is affiliated with the Department of  Computer Science, Heinrich-Heine-University D\"usseldorf, Universit\"atsstra\ss{}e 1, 40225, D\"usseldorf, Germany.} 
\and 
Melanie Schmidt\footnotemark[1]~\thanks{M. Schmidt (\email{mschmidt@hhu.de}) is affiliated with the Department of  Computer Science, Heinrich-Heine-University D\"usseldorf, Universit\"atsstra\ss{}e 1, 40225, D\"usseldorf, Germany.}~\footnotemark[6]
\and Charlotte Zäske\thanks{C. Zäske (\email{charlotte.zaeske@rwth-aachen.de}) is affiliated with the Department of Diagnostic and Interventional Radiology, University Hospital Aachen, Pauwelsstraße 30, 52074, Aachen, Germany} 
\and Liliana L. Caldeira\thanks{L. L. Caldeira (\email{liliana.caldeira@uk-koeln.de}) is affiliated with the Department of Radiology,
University Hospital of Cologne, Kerpener Str. 62, 50937, Cologne, Germany.} ~\thanks{L. L. Caldeira and M. Schmidt have jointly supervised this project.}}

\maketitle

\begin{abstract}
  We consider a missing data problem in the context of automatic segmentation methods for Magnetic Resonance Imaging (MRI) brain scans. Usually, automated MRI scan segmentation is based on multiple scans 
  (e.g., \tonelong, \ttwolong, \tonec, \flair). However, quite often a scan is blurry, missing or otherwise unusable. We investigate the question whether a missing scan can be synthesized. We exemplify that this is in principle possible by synthesizing a \ttwolong scan from a given \tonelong scan.

  Our first aim is to compute a picture that resembles the missing scan closely, measured by \emph{average mean squared error} (MSE). We develop/use several methods for this, including a random baseline approach, a clustering-based method and pixel-to-pixel translation method by Isola et al.~\cite{isola2017} (\pixtopix) which is based on conditional GANs. The lowest MSE is achieved by our clustering-based method. 

  Our second aim is to compare the methods with respect to the effect that using the synthesized scan has on the segmentation process. For this, we use a DeepMedic model trained with the four input scan modalities named above. We replace the \ttwolong scan by the synthesized picture and evaluate the segmentations with respect to the tumor identification, using Dice scores as numerical evaluation. 
  The evaluation shows that the segmentation works well with synthesized scans (in particular, with \pixtopix methods) in many cases.
  \end{abstract}

\section{Introduction}
\label{sec:introduction}

Automated brain tumor segmentation is an active and highly relevant research field~\cite{yan2022, padmapriya2022, ranjbarzadeh2023} with its own annual data challenge called \brats (\emph{Brain Tumor Segmentation Challenge}).
Given a set of MRI scans, the objective of automated tumor segmentation is to find and segment a tumor in the depicted brain automatically. 
Segmentation software (e.g., \deepmedic~\cite{kamnitsas2016}, nnU-Net~\cite{isensee2020}) is not only able to find the exact tumor location, but also to distinguish between different subregions of the tumor, such as edema and necrosis.
These methods~\cite{kamnitsas2016, isensee2020} use a set of four modalities to achieve this purpose, namely \tone (T1-weighted), \ttwo (T2-weighted), \tonec (T1-weighted contrast enhanced) and \flair. 
However, in the clinical routine, it is uncommon that all modalities are available for a single examination.
While it is plausible to create separate networks for each subset of available scans, a more convenient approach is to address the missing data directly and replace the missing scan. Indeed, a recent BraTS challenge~\cite{DBLP:journals/corr/abs-2305-09011} states that \emph{the ability to substitute missing modalities and gain segmentation performance is highly desirable and necessary for the broader adoption of these algorithms in the clinical routine.}\footnote{Note, however, that our work was mostly done before that challenge was posed, so in particular, we are using the BraTS 2019 data set for our experiments.}
We showcase that creating synthesized scans is possible to a certain extent by investigating the following question: 
\begin{center}  
How can we produce a reasonable synthesized \ttwo image from a \tone scan of the same patient?
\end{center}

\begin{figure}[bht]
\centering
\includegraphics{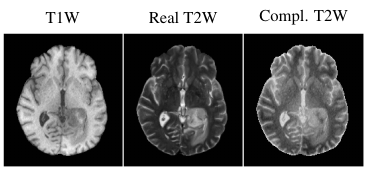}
\caption{A slice of a \tone and a \ttwo scan, and of the  \tone complement (Patient \patient{BraTS19 CBICA BLI 1}).\label{fig:complementary}}
\end{figure}

\cref{fig:complementary} shows a slice of a \tone scan and the corresponding slice of the \ttwo scan of the same patient (these scans are 3D, but we only depict one slice). Since \ttwo looks complementary to \tone, a first idea might be to inverse every voxel, i.e., to transform the voxel intensity $ p_{\tone} $ of \tone 
to $p_{\ttwo} = \min(\tone) + \max(\tone) - p_{\tone}$ where $ \min(\tone) $ and $ \max(\tone) $ are the minimum and maximum values of all voxels in \tone. This transformation is what created the third picture in \cref{fig:complementary}. Our goal is to substantially beat this baseline approach. 

We develop and evaluate two methods. The first one is based on clustering. It uses a training data set to understand the relationship between \tone and \ttwo. 
For a set of given \tone and \ttwo scans, the method learns the intensity spectrum of each tissue in \tone and how it is translated into the corresponding spectrum that the tissue has in \ttwo. This is done by identifying areas of similar color in both pictures via clustering, and  then mapping compatible areas to each other in order to compute a color transformation table. While clustering methods have been applied in medical imaging, the primary focus has been on aiding brain tissue segmentation~\cite{caldeira2009, al2016, mirzaei2018, malathi2018, li2022}. Our \brainclustering approach is conceptually different from other works in the area and might lead to further work on related problems. 
We describe our clustering method in detail in \cref{sec:brainclustering}. The second method is an application of a known tool named \pixtopix~\cite{isola2017, wang2018} that has already been used in a variety of medical applications for image-to-image translation~\cite{yang2020, abdelmotaal2021, haubold2021, bazangani2022}. We use it with slight modifications of the software that we describe in \cref{sec:pixtopix}.
In follow-up work, Raut\etal~\cite{raut2024} uses this same modified version of \pixtopix to generate and evaluate different mappings (e.g., \ttwo to \flair).

Our evaluation is based on a data set based on 460 patients from \brats 19~\cite{brats1, brats2, brats3}. We both evaluate how close the synthesized picture is to the true scan with respect to minimum mean squared error (MSE) as well as how using it affects the segmentation process. For 335 patients, the BraTS 19 data is supplemented with a ground truth segmentation. A radiologist segmented the scans for 15 additional patients to increase the number of cases with ground truth. 110 cases remain that have no ground truth.

\section{\brainclusteringnospace}
\label{sec:brainclustering}
\newcommand{\tissuevar}{{t}}
\newcommand{\colorvarcaps}{{I}}
\newcommand{\colorvar}{{i}}

\begin{figure*}[h!]
    \centering
    \includegraphics[scale=0.8]{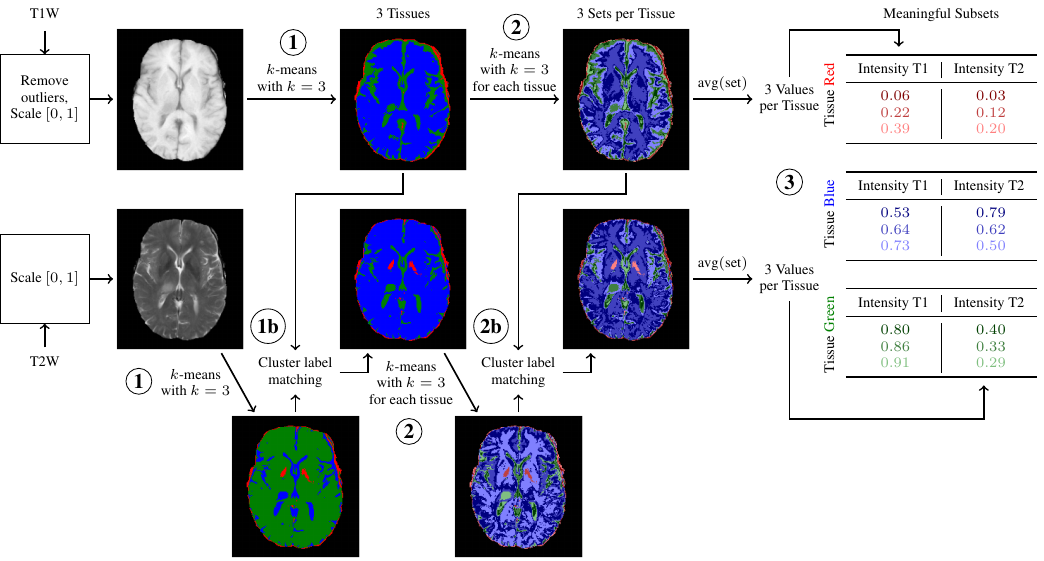}
    \caption{[Best viewed in color] Schematic representation of the \brainclustering process: First, we use \kmeansoned to obtain a macro clustering of the scan according to the desired number of macro clusters. Second, each macro cluster is clustered into micro clusters. Finally, the points belonging to these micro clusters are averaged and stored in one table for each tissue, where there is an entry for each corresponding value of \tone and \ttwo.}
\label{fig:bc_intro}
\end{figure*}

We name our clustering algorithm \brainclustering. It computes a synthesized \ttwo image by applying a chain of several clustering and mapping steps. The clusterings are used to build mapping tables, and these tables are in turn used to generate a synthesized \ttwo. We refer to the process of building these mapping tables as \emph{training}, even though it does not involve a neural network model. 
The following training steps are applied to pairs of given \tone and \ttwo which we process to understand how to construct a \ttwo image from a \tone image. \cref{fig:bc_intro} gives an overview of the process for a single patient, i.e., for one pair of \tone and \ttwo.

\textbf{Training Step 1:} 
Both images are segmented into \emph{macro clusters}. The segmentation is done by solving a one-dimensional $k$-means problem optimally by dynamic programming with the implementation \kmeansoned~\cite{wu1991, gronlund2017, kmeansoned}. In the example, the number of macro clusters $k$ is $3$. Since the macro clusters are supposed to correspond to the different tissues, they should intuitively be equal to the number of different tissues.
However, the process works better if we allow a little more macro clusters to account for varieties of tissues. Empirically, a good range is between three and six clusters.

\textbf{Training Step 1b:} As an intermediate step after finding the macro clusters, we need to identify which cluster represents what tissue. The tissues in \tone can be identified by ordering the clusters according to the average intensity value of their pixels. We could do a similar approach to finding the tissues in \ttwo, but we instead find them by matching them to the found clusters in \tone as described below in Step 2b (cluster label matching).

\textbf{Training Step 2:} Next, we compute a \emph{micro clustering} for each tissue which captures the shadings of the tissues and the relation between the shading in \tone and \ttwo. We again implement this step by optimally solving a one-dimensional $k$-means problem. The number of micro clusters corresponds to the number of different shades that we allow. In \cref{fig:bc_intro}, we use $3$ micro clusters per tissue for visualization purposes, while in practice, we use at least $100$ micro clusters.

\textbf{Training Step 2b:} Now, there are micro clusters for every tissue, both in \tone and in \ttwo. They are small patches of the same shading, and we aim to identify how such a patch in \tone is mapped to \ttwo. For this, we need to match the clusters in \tone and \ttwo. We call this process \emph{cluster label matching}.
The matching is done by first excluding voxels that are only present in one of the scans (since the scans are registered to the same template, these are only a few voxels), and then finding the label that maximizes the number of voxels that the matched clusters have in common. The problem reduces to maximum weighted matching problem, which we solve with the Hungarian method~\cite{jonker1987}.

After this step, we have a macro clustering and every macro clustering is subdivided into small patches, and every small patch in \tone has its specified counterpart in \ttwo.

\textbf{Training Step 3:} The next step is to capture the relationship between the shadings in \tone and \ttwo. The idea behind this is that we assume that it is possible to map the shading of a specific tissue in \tone to the shading of the same tissue in \ttwo. We model this by using a function $ f_\tissuevar $ for every tissue which is supposed to translate intensity values from \tone to intensity values in \ttwo. We later want to use the function $ f_{\tissuevar}(\colorvar_1) = \colorvar_2 $ to predict the \ttwo intensities of a patient whose \ttwo scan is missing.

For every tissue type \tissuevar, we have computed a micro clustering and matched the micro clusters between the two scans. Now we compute the average of the points in each micro cluster of \tone, which we call $ a_1 $. Then, we compute the average for each corresponding micro cluster of \ttwo, which we call $ a_2 $. Finally, we add a new entry $ (a_1, a_2) $ to $ f_{\tissuevar} $, which is the map corresponding to the current macro cluster/tissue type.
It may happen that there already is an entry with key $a_1$ stored in the map. In this case, we compute the average with the moving average formula:
\begin{equation}\label{eq:movingavg}
	f_{\tissuevar}(a_1) \gets f_{\tissuevar}(a_1) + \frac{a_2 - f_{\tissuevar}(a_1)}{\#_\tissuevar(a_1)},
\end{equation}
where $\#_\tissuevar(a_1)$ is the number of times we tried to add an entry with key $a_1$ to the map $f_{\tissuevar}$. Note that consequently we have to keep track of $\#_\tissuevar(a_1)$, i.e., we have to annotate each entry of the map with the corresponding cardinality.

\textbf{Training Step 1-3} are done for all \tone/\ttwo scan pairs available in the training data set, and the values found in Step 3 are always inserted into the same tables. So we get as many tables as we have macro clusters, and up to as many rows as we have micro clusters for \emph{all} training images. Since we process many scans, the full tables get too large, and in reality, we only store a meaningful subset of the rows. When querying for a \ttwo, we can compute missing values by interpolation.

What is left to describe is how we now synthesize images based on our mapping tables. We implemented two options (the second one is faster).

\textbf{Synthesizing \ttwo}:
For a patient whose \ttwo is missing, we preprocess the \tone scan and then cluster it with \kmeansoned to obtain a macro clustering. We load all tables $f_{\tissuevar}$ into the memory which were computed with the same number of macro clusters.
Let $ \tissuevar $ be one of the macro clusters. We consider each point $ p $ belonging to cluster $ \tissuevar $. We find the two rows of $f_{\tissuevar}$ with the closest intensity values to $p$ and interpolate $f_{\tissuevar}(p)$ from these two rows.
The resulting intensities form a synthesized scan. Once it is computed, it is postprocessed using a $ 3\times3 $ median filter~\cite{opencv} to remove \emph{salt-and-pepper} noise which was present in the synthesized images.

\textbf{Synthesizing \ttwo in Search Mode}:
We noticed that having large tables computed by training on many images (e.g., 200) produces noise in the results.
Thus, we propose an additional method that does not precompute a large model before querying, but instead computes a small model every time we want to answer a query, which we call \emph{\search}.
Given a \tone query image, we first search in the training data set for the $ w $ patients whose \tone has the smallest mean squared error to the query \tone (where $ w $ is a small constant, e.g. $ w = 5 $).
Then, we create a small model by performing the training process only on these $ w $ patients, and produce the synthesized \ttwo with this model.
This approach has the downside that we have to create a new model for every single query, thus significantly increasing the query time.
However, since we choose small values for $ w $, a synthesized \ttwo image for one input can be computed in around ten minutes, without the need of computing a model beforehand.

We compare the classic Train\&Test mode and \search mode with $ w \in \{5, 10\} $ and using $ 3, 4, 5 $ and $ 6 $ macro clusters, for a total of 12 different models.

\section{\pixtopix}
\label{sec:pixtopix}
\pixtopix~\cite{isola2017, wang2018} is a general solution to the pixel-to-pixel translation problem.
Given an input image and an output image, this neural network framework is able to learn a relationship between the two.
At inference time, it applies the learned function to create a new, previously unseen image. 
\pixtopix is based on \emph{conditional GANs} (cGAN) ~\cite{mirza2014}.
A cGAN is a neural network framework that simultaneously trains two models, a \emph{generator} $ G $ and a \emph{discriminator} $ D $, which engage in an adversarial process. 
During training, the generator starts with random noise as input, aiming to produce data resembling the training set. The discriminator evaluates the generated data, providing feedback to adjust the generator's parameters. Simultaneously, the discriminator learns to better distinguish between real and generated data, refining its parameters to maximize correct classifications.

The \pixtopix objective function is a fusion between the traditional cGAN loss \cite{mirza2014} and the L1 loss.
We build our objective function by further expanding this concept.
Let $ G $ and $ D $ be any non-linear mapping functions and let $ x, y$ represent our input and our desired output respectively.
The generator is asked to learn the distribution of the data $ y $ from noise $ z $ such that it represents a mapping from a given input $ x $, i.e., $ G: \{x, z\} \rightarrow y $.
The discriminator with mapping $ D({y}) $ outputs the probability that $ {y} $ comes from the training data.
In fact, we train $ D $ to maximize the probability that the real data $ {y} $ is identified as real, while we train $ G $ to maximize the probability that the data generated from noise $ G({z}) $ is detected as real by the discriminator. The latter objective corresponds to minimizing $ 1 - D(G({z})) $. The final form of \cref{eq:pixobj1} below comes from formalizing these high-level objectives with the log-loss principle of multi-class classification problems.
Additionally, Isola\etal~\cite{isola2017} add the term \eqref{eq:pixobj2}, which means that the objective of the discriminator is unchanged, whereas the generator is asked to trick the discriminator while maintaining a small loss. 
Since our objective is to have a low mean squared error in the tumor area, such that tumors can be easily identified, we added~\eqref{eq:pixobj3} to the loss function. $ T $ is a matrix that is zero where there is no tumor and one otherwise, $ \odot $ is the Hadamard product and $ \text{sum}(T) $ is the grand sum of the matrix $ T $, i.e., the number of tumor voxels. 
Our loss function is
\begin{alignat}{2}
&\label{eq:pixobj1} \text{\small$=\expe{\log D({x}, {y})}{{x}, {y}} + \expe{\log(1 - D({x}, G({x}, {z})))}{{x}, {z}}$} \\
&\label{eq:pixobj2} \text{\small$+\lambda \cdot \expe{\lVert {y} - G({x}, {z})\rVert_1}{{x}, {y}, {z}}$}\\
&\label{eq:pixobj3} \text{\small $+\gamma \cdot \expe{\frac{1}{\text{sum}(T)} \cdot \lVert T \odot y - G(T \odot x, T \odot z)\rVert_2^2}{{x}, {y}, {z}}$}.
\end{alignat}
As it is usual in machine learning, the best values for $\lambda$ and $\gamma$ can be found with a hyperparameter optimization.

Additionally, we implement some other modifications:
1) We modify the loading functionality to accept \threed \nifti images.
2) We extend the data augmentation process with an additional library, TorchIO~\cite{perez2020}, which implements useful preprocessing and data augmentation routines for medical imaging.
3) We add the mixed precision training functionality from NVIDIA APEX (available at \url{https://github.com/NVIDIA/apex}). Operations like matrix to matrix multiplication and convolution are then performed in half-precision floating-point format, which results in a speed-up at training time.
4) Since a big portion of our scans is background, the L1 loss is only computed on the actual brain voxels.
5) Recent studies have shown that the transpose convolution operation during upsampling creates checkerboard artifacts~\cite{wojna2019, odena2016}.
We implement the solution of Wojna\etal~\cite{wojna2019}: \emph{linear additive upsampling}. 
This method has been successfully used for the generator of \pixtopixhd~\cite{wang2018} and applied on CT scans~\cite{haubold2021}.
We substitute the transpose convolution operation with a two-factor upsampling followed by a 4-factor reduction of the number of channels, and finally by a $ 3\times3 $ convolution with stride 1. We use bilinear upsampling if the scans are \twod, and trilinear upsampling for the \threed case.

The default generator architecture for \pixtopix is a \unet~\cite{ronneberger2015}, more specifically a \unettwo, which requires an input with side length multiple of $ 256 $.
Given the 3D nature of the scans and the increased memory usage, we opt for a \unetone model with seven downsampling and seven upsampling blocks.
During training, the images are preprocessed to have side length 128 by performing random crops, while during inference they are padded to have side length 256.
Additional augmentation is done by randomly flipping the images.
We also test whether \resnet s~\cite{he2015}, which are used by \pixtopixhd~\cite{wang2018}, are a better choice for our model.
We apply the Adam solver~\cite{kingma2017} with a learning rate of $ 0.0002 $, the momentum parameters $ \beta_1 = 0.5, \beta_2 = 0.999 $ and a batch size of 1.
All other network parameters are the default ones given by \pixtopix.
We train 16 models per generator architecture (\unet\ or \resnet) with the values $ \lambda, \gamma \in \{0, 100, 500, 1000\} $ for a total of 32 different models.

\section{Experimental Evaluation}

We test \brainclustering and \pixtopix. As baseline, we use the complementary approach described in \cref{sec:introduction} and a simple method that we name \randomapproach, which consists of filling all non-background voxels by uniformly distributed random values.

\subsection{Data}
\label{sec:data}
Our collective contains a total of 460 patients from \brats 19~\cite{brats1, brats2, brats3}.
Each patient has four \threed MRI brain scans (\tone, \ttwo, \tonec, \flair) of size $ (240, 240, 155) $ in \nifti (\emph{Neuroimaging Informatics Technology Initiative}) format.
All scans have been preprocessed to have the same voxel size ($ 1\text{mm}^3 $), are skull-stripped and registered to the same anatomical template.
A ground truth segmentation that shows the exact position of the tumor is present for 350 patients have. 
Most of the tumor ground truth comes from \brats, and a portion of the scans has been manually segmented by a radiologist from the University Hospital of Cologne (C.Z.)
and the segmentation has later been converted to the \brats labels.
A detailed representation of the \brats 19 tumor labels is given by Bakas\etal~\cite{brats2}. Each voxel is identified as either background (no tumor), non-enhancing part of the tumor and necrotic core, excess of fluid associated with the tumor (edema), or contrast enhanced part of the tumor.

We divide the data into a training set, a validation set, and three test sets (\cref{tab:datasets}).
The training set is used to train both approaches, the validation set to optimize the hyperparameters, and the three test sets to evaluate the approaches. We evaluate \testbrats and \testuk separately due to the different sources of ground truth. 

\begin{table}
\centering
\caption{Our split into data sets of the \brats 19 data.}
\begin{tabular}{ c  c  c }
\toprule
Data set & N. Patients & Segmentation \\ 
 & N. Patients & Ground Truth\\ 
\midrule
\train & 270 & \brats 19 \\
\validation & 30 & \brats 19 \\
\testbrats & 35 & \brats 19 \\
\testuk & 15 & anonymous \\
\nogtdataset & 110 & / \\
\bottomrule
\end{tabular}
\label{tab:datasets}
\end{table}

\subsection{Code and System}
\brainclustering is written in C++17 and ported to Python with Boost.Python~\cite{boostpython}, while our \pixtopix implementation is written in Python3.8 and uses the machine learning framework PyTorch~\cite{paszke2019}. Our implementations are available at 
\url{https://github.com/giuliabaldini/brainclustering} and \url{https://github.com/giuliabaldini/Pix2PixNIfTI}.
The \brainclustering scans are produced using a computer cluster with 36 nodes and two Intel Xeon CPU E5-2690 v2 CPUs per node with 10 cores each and 128 GB of RAM.
We compute the \pixtopix scans, the segmentation and the evaluation results on a machine with eight 32 GB NVIDIA Tesla V100-SXM2. 

\subsection{Evaluation Methods}
\label{sec:seg}

We evaluate: 

\emph{a) Visual Similarity to Real Scans,} evaluated by computing 
\emph{Average Mean Squared Errors} (MSEs) to the original scans.
We compute the MSE between the real \ttwo and each synthesized \ttwo in the brain mask voxels and, if the tumor ground truth is available, we additionally compute the MSE restricted to the tumor area. Finally, we normalize all values to $[0,1]$.

\emph{b) Usefulness for Segmentation:}
We use a \deepmedic~\cite{kamnitsas2016} model to evaluate the usefulness of the synthesized images for segmentation. The model was computed on the \train data set with all four MRI scans and the corresponding tumor ground truth as input. The hyper parameters were chosen according to a clinical evaluation on \deepmedic~\cite{perkuhn2018}. The model is trained to predict three tumor classes: necrotic/non-enhancing tumor, edema and contrast enhanced tumor.

For the evaluation, we compute the segmentation with the \deepmedic model on the test data sets using the synthesized \ttwo (instead of the real \ttwo scan) and the real \tone, \tonec and \flair.
We use Dice scores~\cite{bertels2019} and \hausdorff~\cite{plastimatch} (HD95) as evaluation metrics. 
For the case of \testbrats and \testuk, we compute the Dice score and the HD95 between the segmentation produced with the synthesized images and the ground truth segmentation. For \nogtdataset, these values are computed in comparison to the segmentation produced with the real \ttwo (instead of the ground truth segmentation). 
Following the \brats evaluation~\cite{brats1, brats2, brats3}, we assess the tumor classes based on three categories: whole tumor (necrotic/non-enhancing tumor, edema and contrast enhanced tumor), tumor core (necrotic/non-enhancing tumor and contrast enhanced tumor), and active tumor (contrast-enhanced tumor) and then compute the average of the three cases.

\section{Results}
\label{sec:results}
\begin{figure}
\centering
\includegraphics[scale=0.9]{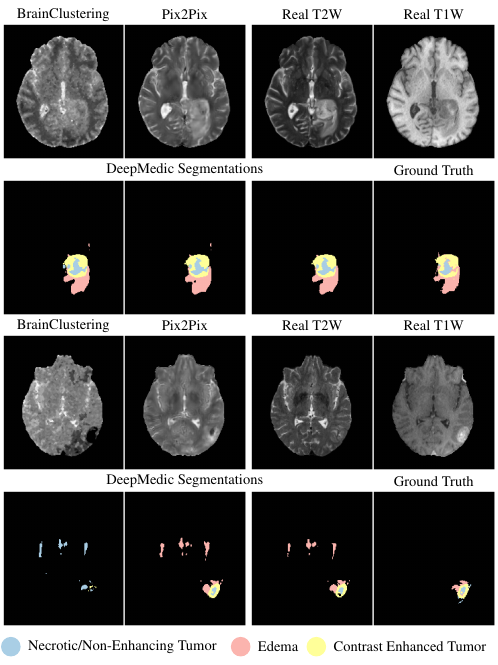}
\caption{
%Comparison of the segmentation results. 
The segmentations in the second and in the fourth row have been created using the respective \ttwo and the three original \tone, \tonec, \flair. The rightmost image in the \enquote{Real T1W} column shows the real \tone image and the ground truth segmentation.
}
\label{fig:comp}
\end{figure}

\cref{fig:comp} up to \cref{fig:testuk2} give a visual impression of the resulting synthesized images for the \testuk data set. 
In \cref{fig:comp}, we present two synthesized images produced by \brainclustering and \pixtopix for two patients.
The second and fourth row show the segmentation achieved by using the respective image as \ttwo. The rightmost image in the \enquote{Real T1W} column shows the ground truth image and segmentation.
\cref{fig:testuk1} and \cref{fig:testuk2} present the middle transversal slice for each MRI in \testuk with the segmentation superimposed on the scans.

\subsection{Run Time}
\brainclustering \traintest takes $ 2.2 $ hours on average to train on 270 patients, and testing takes around $ 9 $ seconds per scan.
By comparison, \search 5 and \search 10 take respectively $ 2.5 $ minutes and $ 4.8 $ minutes per patient.
The \pixtopix \resnet\ training takes around $ 9 $ hours, while the \pixtopix \unet\ training takes around $ 3 $ hours. 
The testing process takes around $ 7 $ seconds per scan.
The baseline methods can be computed in negligible time since only one pass over \tone is required for \complementary and \randomapproach can be computed without looking at the input data.

\subsection{Mean Squared Error Evaluation}
\begin{figure*}[t]
    \centering
    \includegraphics[scale=0.89]{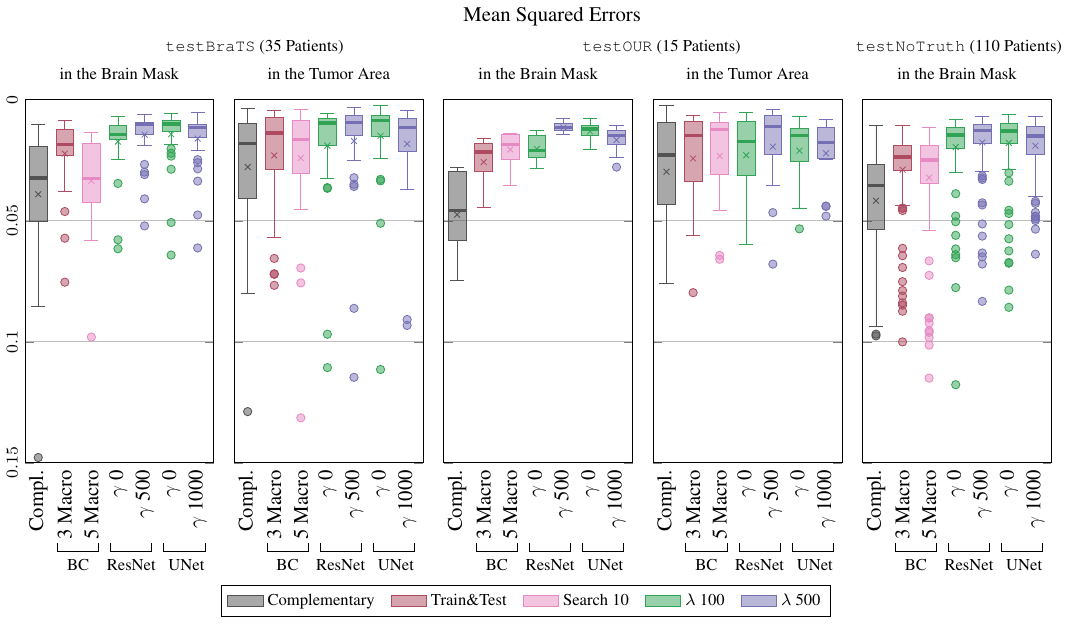}
    \caption{[Best viewed in color] Comparison of different approaches with respect to MSE in the brain mask and (if available) in the tumor areas. The scores represent the distribution of the patients. A lower value indicates a better score. The crosses represent the averages, the thick bars are the medians and the dots are the outliers.}
	\label{fig:test_mse}
\end{figure*}

In \cref{fig:test_mse}, we present the MSE results for the \testbrats, \testuk and \nogtdataset data sets. To improve clarity, we have selected two \brainclustering, two \pixtopix \resnet\ and two \pixtopix \unet\ models, which are those that performed best on the validation data set in the respective category.
\randomapproach has been excluded from the figure because its MSEs are too large (averaging at $0.37$ on the brain mask and $0.11$ on the brain tumor area).  
The second baseline, \complementary evaluates to reasonable MSEs, e.g., it has $0.04$ on average in the brain mask and $0.03$ in the tumor area on \testbrats. All our methods improve upon these values as can be seen in \cref{fig:test_mse}. The scans generated by \pixtopix have a better MSE than \brainclustering in the brain mask area. In the tumor area, the average MSE of both methods is comparable, and it is overall better, although the median MSE is still below 0.02. The best results are produced by \pixtopix \resnet\ with $\lambda=500$ and $\gamma =500$, but \pixtopix \unet\ with $\lambda=100$ and $\gamma=0$ also produces good results. Notably, \brainclustering with \search with $10$ patients and $5$ clusters achieves the best median MSE in the tumor area among all methods we tested. We conclude that it is possible to compute synthesized scans that closely resemble the true scan.

\subsection{Segmentation Evaluation}
\begin{figure*}[t]
    \centering
    \includegraphics[scale=0.89]{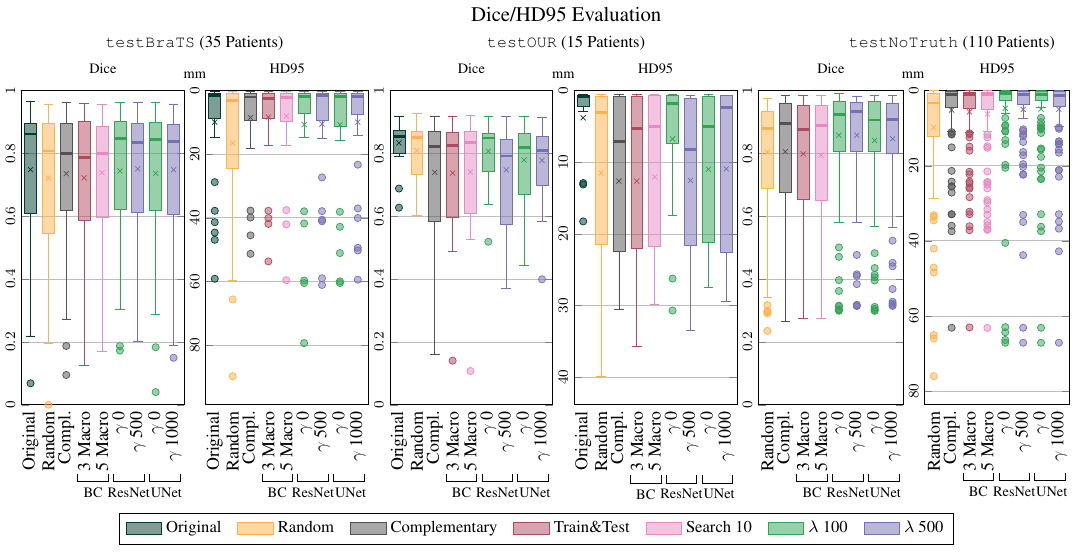}
    \caption{[Best viewed in color] Comparison with respect to the true tumor segmentation (for \testbrats and \testuk) and the original tumor segmentation (for \testnotruth) using Dice and undirected $ 95^{\text{th}} $ Hausdorff distance (HD95). For \testbrats and \testuk, we also include the segmentation which is produced by using the real \ttwo as \emph{Original} to give a positive baseline as to what scores are achievable if \ttwo was reproduced perfectly. We compute the Dice score and the undirected 95\% percent Hausdorff distance on three different regions and average the scores. The box plots represent the distribution of the patients. For the Dice score a higher value indicates a better value whereas for the Hausdorff distance the opposite holds. The crosses represent the averages, whereas the thick bars are the means and the dots are the outliers.}
	\label{fig:test_rank}
\end{figure*}

\cref{fig:test_rank} shows the segmentation results for the same models. 
All our methods perform better than \complementary, and \pixtopix performs better than \brainclustering. 
For \pixtopix, the \resnet\ models are in most cases better than the \unet\ ones. 
The segmentations with the original \ttwo obtain an average Dice score of 0.833 on \testuk and of 0.748 on \testbrats, and an average of 0.774 for both data sets together. 
The overall best performing \pixtopix approach is with $\lambda=100$ and $\gamma=0$, with an average Dice of 0.807 on \testuk and of 0.744 on \testbrats, and an average of 0.763 for both data sets together. 
For \brainclustering, the best method is \search with 10 scans, 5 macro clusters and 100 micro clusters, with an average Dice of 0.741 on \testuk and of 0.739 on \testbrats, and an average of 0.739 for both data sets together. 
This method achieves an average HD95 of 12.047 on \testuk, of 8.021 on \testbrats and of 9.229 for both data sets together.
On \nogtdataset, the best \pixtopix model achieved a Dice score of 0.858 and an HD95 of 4.706 and the best \brainclustering model achieved a Dice score of 0.794 and an HD95 of 6.213.

\twocolumn

\begin{figure}
\centering
\includegraphics[scale=0.9]{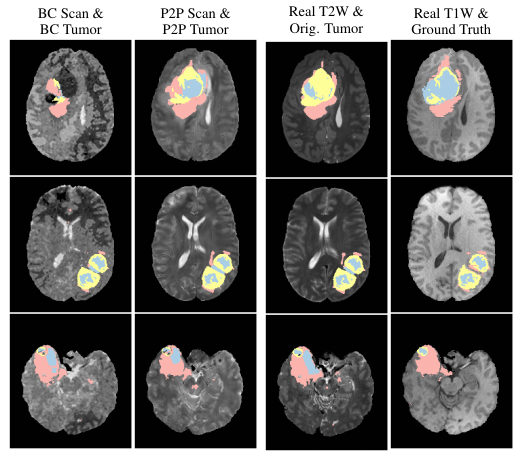}
\caption{%
\brainclustering, \pixtopix and original \ttwo scans and corresponding superimposed segmentation generated with \deepmedic (together with the real \tone, \tonec and \flair), shown for all cases from the \testuk data set. In the last column the ground truth segmentation is superimposed on the \tone scan.
}
\label{fig:testuk1}
\end{figure}
\begin{figure}
\centering
\includegraphics{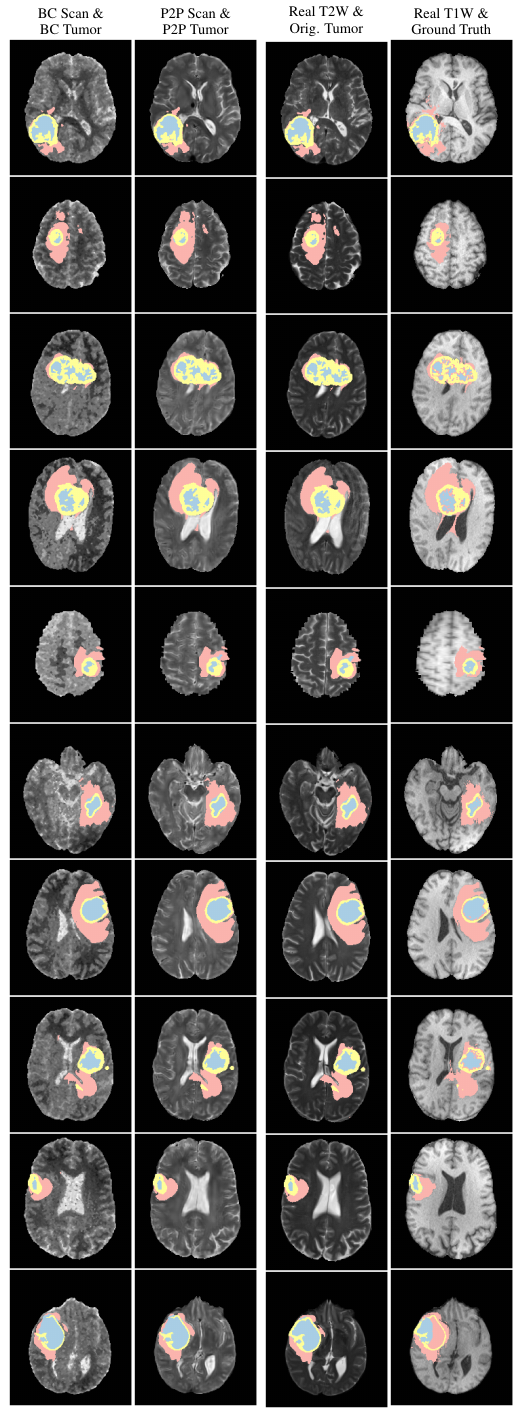}
\caption{Continuation of \autoref{fig:testuk1}.}
\label{fig:testuk2}
\end{figure}
\onecolumn
\clearpage

\clearpage

So note that for the data set with ground truth (\testuk and \testbrats), the best performing \pixtopix method leads to an average Dice score of $0.763$ compared to an average score of $0.774$ when using the real T2-weighted, so it is very close (\brainclustering performs a bit worse, achieving an average score of $0.739$).
\begin{table}[thb]
\caption{Mean and standard deviation of (Method-Dice - Original-Dice) and (Original-HD95 - Method-HD95). Negative values indicate a decrease in quality.
\label{tab:method_distribution}}
\centering
\small 
\setlength{\tabcolsep}{2pt}
% There are more digits available. Melanie removed them due to space constraints. See TMI submission.
\begin{tabular}{ c  c  c  }
\toprule
Approach & \testbrats & \testuk  \\
\midrule
\multicolumn{3}{c}{Average Change in Dice Score}\\
\midrule
\randomapproach    & $ -0.0267 \pm 0.0839 $ & $ -0.0242 \pm 0.0693 $  \\
\complementary     & $ -0.0129 \pm 0.0904 $ & $ -0.0936 \pm 0.1729 $ \\
BC Train/3 Macro   & $ -0.0258 \pm 0.0738 $ & $ -0.0960 \pm 0.1830 $ \\
BC-S 10/5 Macro    & $ -0.0097 \pm 0.1002 $ & $ -0.0920 \pm 0.1862 $ \\
{\resnet\ 100/0}   & $ -0.0038 \pm 0.0561 $ & $ -0.0268 \pm 0.0757 $ \\
{\resnet\ 500/500} & $ +0.0024 \pm 0.0584 $ & $ -0.0857 \pm 0.1271 $ \\
{\unet\ 100/0}     & $ -0.0115 \pm 0.0579 $ & $ -0.0541 \pm 0.0927 $ \\
{\unet\ 500/1000}  & $ -0.0003 \pm 0.0546 $ & $ -0.0558 \pm 0.1024 $ \\
\midrule
\multicolumn{3}{c}{Average Change in 95\% Hausdorff Distance}\\
\midrule
\randomapproach    & $ -6.5477 \pm 12.7633 $ & $-7.7170\pm  10.9374  $  \\
\complementary     & $ +1.4568 \pm 11.1163 $ & $-8.8377\pm 10.3483   $  \\
BC Train/3 Macro   & $ +1.6263 \pm 11.5007 $ & $-8.8494 \pm  11.0154$  \\
BC-S 10/5 Macro    & $ +1.9030 \pm 10.5994 $ & $-8.2579 \pm 10.1501  $  \\
{\resnet\ 100/0}   & $ -0.7786 \pm 7.4124 $ & $-2.9560 \pm  7.9400  $  \\
{\resnet\ 500/500} & $ -0.5746 \pm 5.7906 $ & $-8.7538 \pm 9.7339 $  \\
{\unet\ 100/0}     & $ -0.7717 \pm 5.4584 $ & $-7.1996 \pm 9.1900 $  \\
{\unet\ 500/1000}  & $ +0.0282 \pm 4.4942 $ & $-7.1382 \pm 10.5381$  \\
\end{tabular}
\end{table}

As an additional overview, \cref{tab:method_distribution} shows the difference in the metrics when replacing the real \ttwo by a synthesized \ttwo scan. %affects the segmentation. 
The \pixtopix \resnet\ approach with $\lambda=100$ and $\gamma=0$ decreases the Dice score by less than $0.01$ on \testbrats, and by less than $0.03$ on \testuk. 
The average Dice score of the original segmentations on \testbrats and \testuk are $0.748$ and $0.833$, so the average decreases in Dice score are below 1\% and 4\% of the original score.

An interesting byproduct of our work is that we observe that substituting the \ttwo image with random noise also produced acceptable results. This is a common practice that gets some justification from our evaluation. However, the best tested methods outperform this workaround.

\subsection{Concluding remarks}
A general limitation of trying to reproduce a missing scan is that if something is only visible in the missing scan, then obviously it cannot be synthesized. 
However, our study shows that in many cases, enough information is contained in \tone and a useful synthesized \ttwo can be computed. The \brainclustering pictures are sometimes noisy but achieve a low average MSE. The \pixtopix pictures are a bit too light, but capture the structure very well. The segmentation is often only mildly or not at all affected by replacing the real \ttwo by synthesized ones. 
The proposed methods can thus aid already trained segmentation networks whenever the \ttwo modality is missing.

%\bibliographystyle{alpha}
%\bibliography{bibliography.bib}
\printbibliography

\end{document}